# Different regimes of Purcell Effect in Disordered Photonic Crystals

K.M. Morozov[a], A.R. Gubaydullin[a, b], K.A. Ivanov[b], G.Pozina[d] and M.A. Kaliteevski [a, b, c]

[a] St. Petersburg National Research Academic University, Russian Academy of Sciences, St. Petersburg, 194021 Russia
[b] ITMO University, St. Petersburg, 197101 Russia
[c] Ioffe Institute, Russian Academy of Sciences, St. Petersburg, 194021 Russia
[d] Department of Physics, Chemistry and Biology, Linköping University, 58183, Linköping, Sweden.

**Abstract**

We demonstrate that disorder in photonic crystals could lead to pronounced modification of spontaneous emission rate in the frequency region corresponding to the photonic band gap (PBG). Depending on the amount of disorder, two different regimes of the Purcell effect occurs. For the moderate disorder, an enhancement of spontaneous emission occurs at the edge of PBG due to modification of the properties of the edge state. This effect is responsible for recently observed mirrorless lasing in photonic crystals at the edge of PBG. When the level of disorder increases, the spontaneous emission rate enhances within the PBG due to the appearance of the high quality factor states. This effect is likely responsible for a superlinear dependence of emissions on the pumping observed in synthetic opals.

**Introduction**

Interplay of the Bragg interference and random scattering of light in disordered photonic structures [1–3] gives rise to a wide range of fascinating optical phenomena, such as Anderson localization of light [4,5] peculiar transport [6–8] and emission properties [9–11]. Historically, the behaviour of the wave in disordered media has been considered for electrons in disordered solids [12,13], where useful theoretical approaches (such as scaling theory of localization [14–16]) were developed and interesting effects (such as Mott transition [17] and Anderson localization [18]) were predicted.

Despite some similarities with behaviour of electrons in disordered solids and photons in the disordered photonic crystals (PC), there are important differences between the two cases. Firstly, for electrons, efficient electron – phonon and electron-electron interactions and inelastic scattering on atomic potential fluctuations substantially reduce the electron coherence length, making it in most cases much smaller than the size of the experimental sample. At the same time, photons do not interact with each other in "linear" materials, and the coherent length for photons exceeds a sample size in non-absorbing materials. Another important difference originates from the differences between wave equations for electrons and for electromagnetic fields. The Schrödinger equation for electron is read as:

$$-\frac{\hbar^2}{2m}\nabla^2\psi(\mathbf{r}) = (U(\mathbf{r}) - V(\mathbf{r}))\psi(\mathbf{r}) \quad (1)$$

It possess' a potential term, which could be negative that automatically leads to the formation of localized electron states. For the electromagnetic wave equation, we have:

$$\nabla \times \nabla \times \mathbf{E}(\mathbf{r}) = \varepsilon(\mathbf{r})\frac{\omega^2}{c^2}\mathbf{E}(\mathbf{r}) \quad (2)$$

where the corresponding term is always positive, and the light localization could only be achieved within the interference [19,20]. It was noticed that such localization can easily occur at the edges of the photonic band gap (PBG) [21]. Previously, it was confirmed that there are different regimes of localization of light in disordered photonic crystals [22,23] and the localization of light can be achieved more easily in the band edge regions. In disordered PC, in the frequency region corresponding to PBG, the gap is filled with localized states, and for such states the local magnitude of the electric field could be very high [23]. Such localized states act as a Fabry-Perot resonances and lead (in contrast to the case of electron in disordered solid) to increasing transmission through the structure [20,22].

Strong spatial variation of the electromagnetic field for the localized photonic states in disordered photonic crystals may also lead to a pronounced modification of the spontaneous emission rate (Purcell effect) [9,11,19,20,24–29]. Thus, our work is aimed at detailed analysis of influence of disorder on the probability of spontaneous emission in photonic crystals.

The probability of spontaneous emission can be estimated by a number of methods, for example by the Green function approach [30] or by analysis of spatial variation electromagnetic field in spatially inhomogeneous structures [31]. The approach developed in [31] requires less computational power (that is essential for our tasks in view of necessity of statistical analysis of the effect) but it works only for the structures possessing a centre of symmetry. Here, for the analysis of Purcell effect in disordered PC, we will use generalization of the approach developed in [30], referred as S-quantization [32–34] (see Appendix 1).

**The model**
As the model, we consider a one-dimensional periodic structure consisting of a sequence of pairs of layers A and B with the same thickness $D/2$, whose refractive indices are

$$n_{A,B}^{(I)} = n_0 \pm g, \qquad (3)$$

where $g$ is the modulation of the refractive index and $n_0$ is the average refractive index. An example of the refractive index profile for such a structure with parameters $n_0 = 2.0$; $g = 0.025$ is illustrated in Figure 1 by the solid line. Propagation of waves in a one-dimensional structure is conventionally described by the transfer matrix method, which provides the dispersion equation for photons in such a structure. For infinite structures with period D the dispersion equation reads:

$$\cos(KD) = tr[T(\omega)], \qquad (4)$$

where $T$ is the transfer matrix for one period. Solution of this equation provides PBG when the Bloch wave vector $K$ becomes imaginary. The center of the first PBG is at $\omega_0 = \pi c/(n_0 D)$, the relative width of the gap is $\Delta\omega/\omega_0 \approx 4g/(\pi n_0)$ [35]. In the case of a structure of finite size, the mode spectrum is discrete and the eigenfrequencies can be obtained by setting outgoing wave boundary conditions. Formation of PBG defined a wide utilization of photonic crystals with the purpose of controlling spontaneous emission [36,37].

For study of properties of disordered structures we introduce random fluctuations of the refractive indices: for each pair of layers A and B in the unit cell, the refractive indices are defined by the following formula:

$$n_{A,B}^{(D)} = n_0 \pm g + n_0 \delta P, \qquad (5)$$

where $P$ takes random values in the interval from -0.5 to 0.5, and $\delta$ specifies the level of disorder for a particular structure. We deliberately use top-hat distribution given by (9) (not Gaussian or Cauchy distribution) in order to exclude the small possibility of an appearance of the elements of the systems, which themselves can serve as localization centres for light. An example of the refractive index profile for a disordered structure is illustrated in Figure 1 by the dashed line. Disorder can also be introduced by varying the layer thicknesses, which also leads to fluctuations of the optical length. Previous study has shown a similar impact of those two types of the optical length fluctuations on the localization properties. Therefore, to avoid cumbersomeness of simulated results we focus only on the disorder of the refractive index.

For the model considered in the case of zero disorder, PBG appears at frequency

$$\omega_0 = \pi c/(n_0 D), \qquad (6)$$

and a relative PBG width reads

$$\Delta\omega/\omega_0 = 4g/(\pi n_0) \qquad (7)$$

Hereafter we will use the average refractive index $n_0 = 2.0$ and modulation $g = 0.025$, providing relative width of the PBG $\Delta\omega/\omega_0 \approx 0.016$.

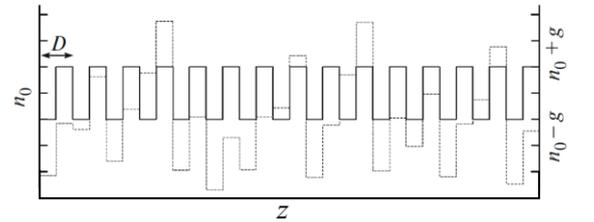

Figure 1. Parameters of the structure. Refractive-index profiles in the ideal and disordered structures are shown by solid and dashed lines, respectively.

**Results and discussion**
In the presence of disorder, exponential tails of the density of states $\rho$, characterized by penetration depth $\Omega$, appear in PBG:

$$\rho \sim \exp\bigl((\omega - \omega_e)/\Omega\bigr)^2 \qquad (8)$$

Penetration depth $\Omega$ relates to disorder parameter $\delta$ as [35]

$$\Omega = \delta\sqrt{\pi/4}\sqrt{\Delta\omega\omega_0} \qquad (9)$$

When penetration depth becomes equal to the half-width of PBG, the localized states can appear in the PBG that defines threshold level of disorder:

$$\delta_{th} = \sqrt{2/\pi}\sqrt{\Delta\omega/\omega_0} \quad (10)$$

and for the parameters used for the modelling $\delta_{th} \approx 0.75$. For the values of the disorder parameter above $\delta = 2\delta_{th} \approx 0.15$ the effects, related to an existence of PBG disappear and the system can be considered completely disordered. In line with [20,22] we will limit our consideration by the one-dimensional case: we assume that an electromagnetic field and dielectric constant are the functions of only one variable *z*, and light propagates along the *z* axis, which, despite a simplicity of modelling can depict important physical effects.

Figure 2 shows examples of the Purcell coefficient dependence on the frequency of light and on the dipole position in the sample (in the following, we will call this dependence as the pattern of the Purcell coefficient) for various values of the disorder parameter $\delta$. Figures 2a shows the pattern of the Purcell coefficient for the ideal structure ( $\delta = 0$ ) of the finite size corresponding to 200 periods. It can be seen that the Purcell coefficient is reduced within the frequency region corresponding to PBG and such a reduction is more pronounced for the central part than for the areas near boundaries of the sample. In contrast, the Purcell coefficient is increased at the edges of PBG, especially at the centre of the sample where it becomes as large as ten. The reason for such enhancement is the formation of edge states [20,38], and such states are responsible for lasing in distributed feedback (DFB) lasers [39,40], and are the most probable reason for mirrorless lasing in photonic crystals observed recently [9]. Similar effects were first predicted [41] and explained in terms of enhancement of the group velocity at the edges of PBG. However, for the disordered system, such concepts as dispersion relation and group velocity lose their meaning, therefore, the approach used in this work is more appropriate. Despite the fact that for the edge states, the field decays slower than exponentially [20], edge states are considered to satisfy Thouless criterion of localization. When the photonic crystal become weakly disordered, the frequencies of edge states fluctuate and PBG lose the shape defined by the ideal structure, as shown in figure 2b. For the states, shifted into PBG, the maximal magnitude of the electric field (and consequently, the Purcell coefficient) increases. For the localized states shifted deep into PBG (shown in figures 2cd), the spatial profile of the field demonstrates an exponential-like decay, maximal values of the field in the mode and the Purcell coefficient increases.

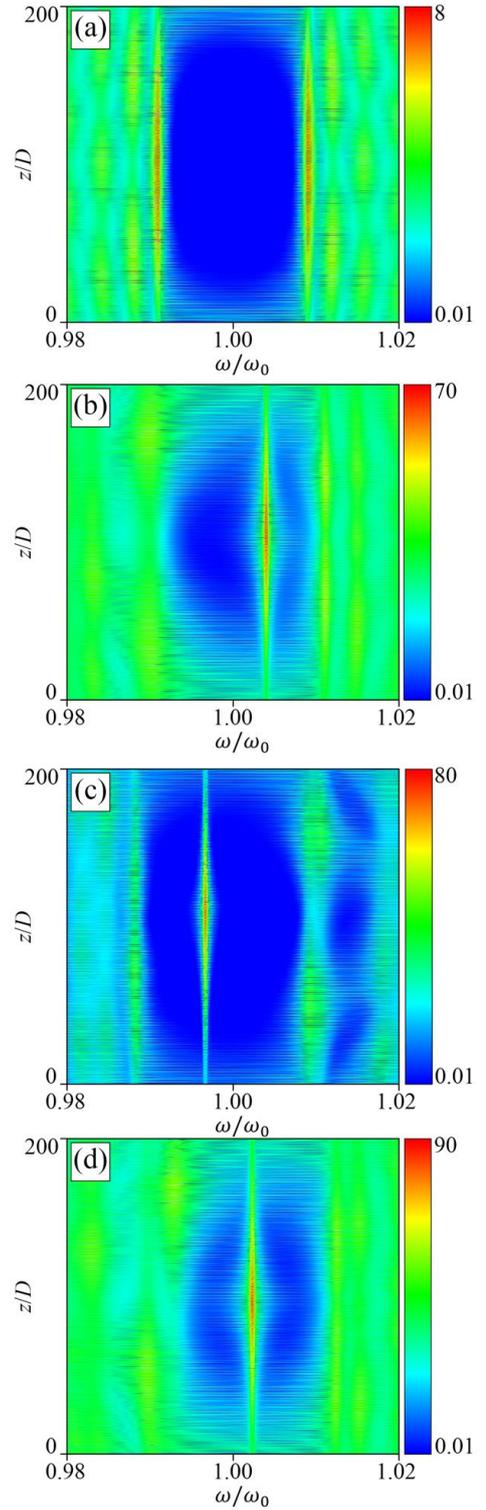

Figure 2. Dependence of the modal Purcell factor on the frequency and position of the dipole source calculated for disordered structure, which provides the maximum value of the modal Purcell factor, obtained in the ensemble of $10^4$ structures. The disorder fluctuation parameter δ=0.07 (b), 0.1 (c), 0.15 (d), and the ideal structure δ=0 (a).

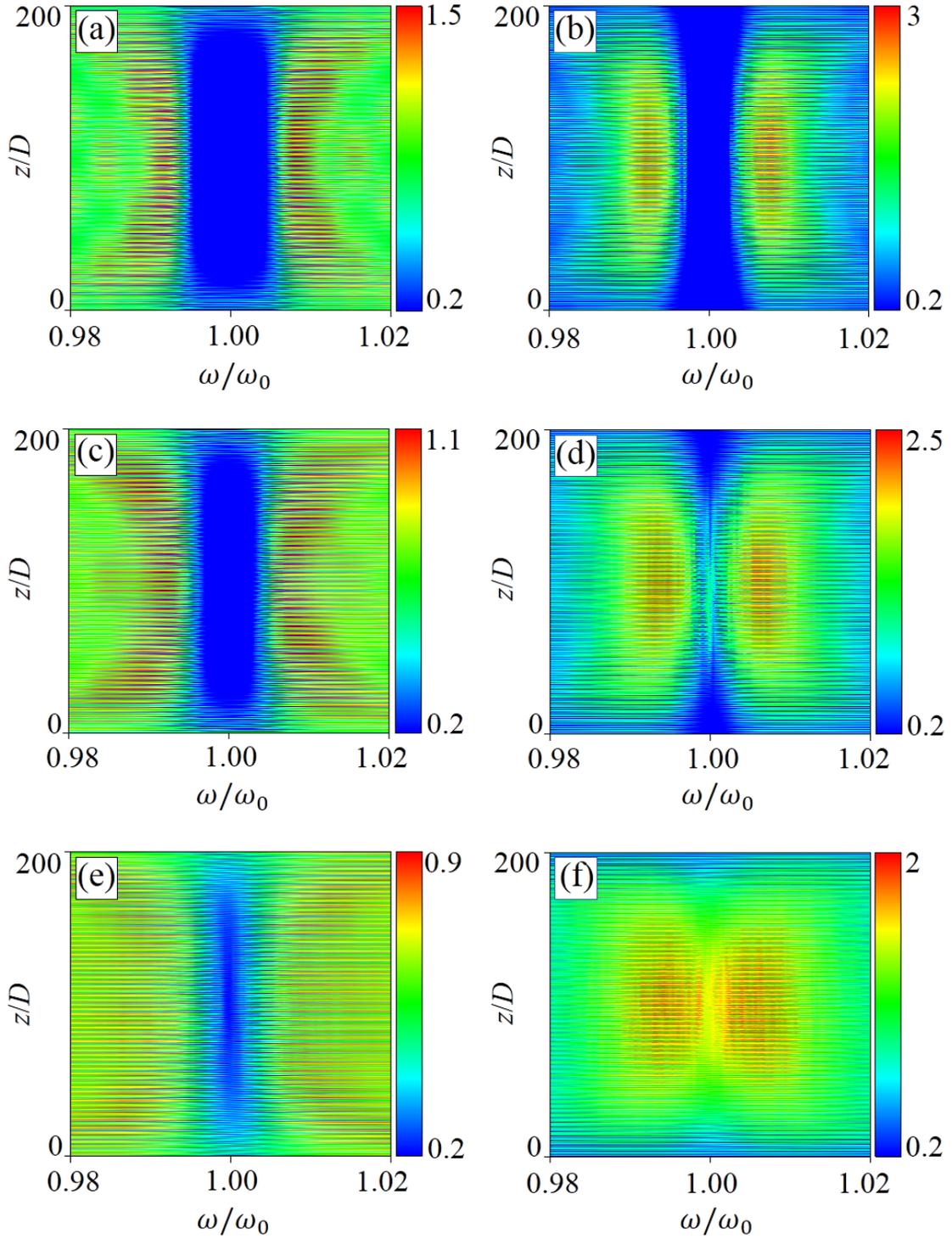

Figure 3. On the left: dependence of the modal Purcell factor on the frequency and position of the emitter placed inside the disordered structure, averaged over an ensemble of $10^4$ structures with δ=0.07 (a) 0.1 (c) and 0.15 (e). On the right: (b, d, f) show the dependence of the standard deviation σ corresponding of the modal Purcell factor, on the frequency and the dipole position. The value of the disorder parameter is (a, b) δ=0.07; (c,d); 0.1; (e,f) 0.15.

When photonic crystals are disordered and the disorder parameter exceeds a threshold value defined by eq. (11), the probability of the appearance of localized states within PBG become noticeable and for the localized state, the local magnitude of the electric field of the mode, as well as the Purcell coefficient could be as high as $10^2$ as shown in figure 2b-d. Note, that for localized states that appear within PBG illustrated in figures 2b-d, the Purcell enhancement of the spontaneous emission rate is an order of magnitude greater than for the edge state.

The modes that appeared within PBG could lead to a lasing in the structure. If the mode is in the vicinity of the PBG centre, the structure works similarly to a vertical cavity surface emitting laser [42], but if the frequency of the mode is close to edges of PBG, the lasing mechanism could be similar to those in DFB lasers [39–41].

For deeper understanding of physics behind the Purcell effect in disordered photonic crystals one should analyse properties of the system "on-average". Figure 3 shows the pattern of the Purcell coefficient and its average over $10^4$ configurations. It can be seen that when the disorder parameter equals its threshold value $\delta = 0.07$, there are two ranges related to a pronounced area of the reduced Purcell coefficient corresponding to the photonic band gap and to a widened area of enhanced emission corresponding to the edge state, respectively. Note that the threshold value of disorder, the mean Purcell factor and its standard deviation becomes equal for the centre of PBG and for edges of PBG, as illustrated in figure 4 where dependencies of the average Purcell coefficient and its standard deviation (also averaged on the position within the structure) are shown. Below the threshold the mean value of the Purcell coefficient is larger than its standard deviation and the properties of the system are similar to properties of ideal photonic crystals. Above the threshold the standard deviation of the modal Purcell coefficient exceeds its mean value and the system should be treated as chaotic. Figures 3 c-f illustrates the evolution of the pattern of the Purcell coefficient with increasing disorder parameter $\delta$. It can be seen, that though the disorder increase leads to the shrinking of PBG for the Purcell coefficient, PBG in the pattern of the Purcell coefficient still remains noticeable for both values of disorder $\delta = 0.1$ and $\delta = 0.15$. At the same time, in the pattern of standard deviation of Purcell coefficient, PBG disappears when $\delta$ reaches the value 0.1. It is interesting to note the differences between the dependencies of $<F>$ and $\sigma$ on the disorder parameter $\delta$ at the centre of PBG and at its edges. For the centre of PBG, both $<F>$ and $\sigma$ demonstrate a monotonic increase with increasing delta, and for disorder values $\delta$ in the threshold area, $\sigma$ could exceed $<F>$ by a factor of ten. In contrast, at the edges of PBG, increasing disorder leads to a monotonic decrease of the Purcell coefficient, and its standard deviation demonstrates a non-monotonic behaviour: first it quickly rises and reaches a maximum at $\delta = 0.02$ and then slowly decreases with an increasing disorder parameter; the deviation $\sigma$ exceeds the mean value $\langle F \rangle$ less than by factor of two.

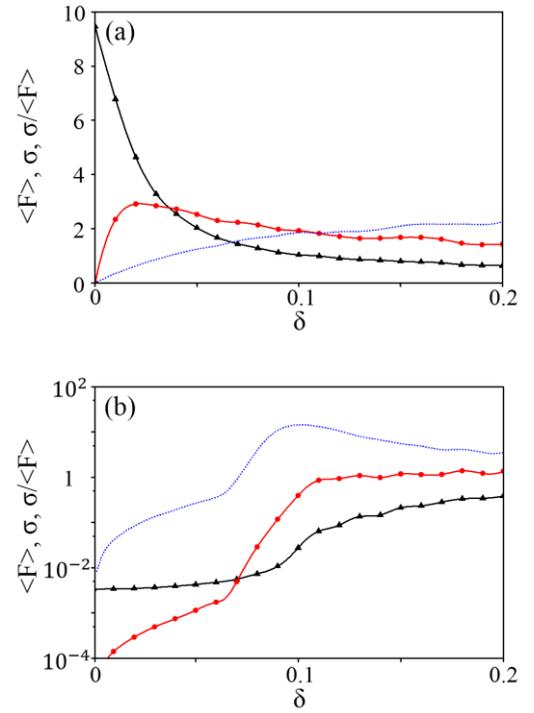

Figure 4. Dependence of mean Purcell coefficient $\langle F \rangle$ (black triangles), its standard deviation $\sigma$ (red circles) and the ratio $\sigma/\langle F \rangle$ (blue dotted) on the value of the disorder parameter $\delta$ for (b) the center of PBG $\omega = \omega_0$ and (a) for the frequency, corresponding to the edge state $\omega = 1.00907\omega_0$.

It is useful to analyse the behaviour of the ratio $\sigma/\langle F \rangle$, which can be considered as a "measure of chaos" in the emission properties of the system. At the edges of PBG (see figure 4a), the ratio $\sigma/\langle F \rangle$ grows monotonically with an increasing disorder $\delta$. At the centre of PBG, the $\sigma/\langle F \rangle$ demonstrates a counter-intuitive result: the dependence of measures of chaos in emission properties as a function of the amount of disorder shows a non-monotonic behaviour: there is a maximum of $\sigma/\langle F \rangle$ for the disorder $\delta = 0.1$, slightly above $\delta_{th}$. Such peculiar behaviour can be explained by an interplay of order (represented by the existence of large pieces of the structure, which are almost periodic and provide Bragg reflection) and disorder (which provides a substantial phase shift between the waves coupled to ordered pieces of the structure).

The pattern of the ratio $\sigma/\langle F \rangle$ is shown in figure 5. It can be seen, that for moderate disorder $\delta = 0.07$ (figure 5a) there are two areas of "increased chaos" and the area of "reduced chaos" corresponding to PBG. The dependence of $\sigma/\langle F \rangle$, $\langle F \rangle$, and $\sigma$ averaged over the position of the emitter in the sample, is shown in figure 5b. It can be seen that for $\delta = 0.07$ there are two peaks near the edges of PBG, and the value of $\sigma/\langle F \rangle$ could be as high as ten. It means that there is a substantial probability of appearance of the state characterized by a high Purcell factor, which exceeds the Purcell factor of the edge states for the ideal structure. For the disorder parameter $\delta = 0.1$ (see fig. 5cd), the two peaks in the dependence $\sigma/\langle F \rangle$ vs $\omega$ at the edges of PBG are replaced by one sharp peak at the centre of PBG, which manifests an establishment of the new regime in the Purcell effect. In this regime, the centre of PBG becomes the most chaotic frequency region, where one can expect an appearance of the states characterized by a high Purcell coefficient. A subsequent increase of the disorder parameter $\delta$ to the value of 0.15 leads to the increase of both $\sigma$ and $\langle F \rangle$, however, $\langle F \rangle$ grows faster, and the peak of the pattern of $\sigma/\langle F \rangle$ become smoothed.

**Conclusion**

Modification of spontaneous emission rate (Purcell effect) in disordered photonic crystals was studied using S-quantization formalism. Calculated dependencies of the Purcell coefficient on a frequency and position of the emitter in photonic crystals shows that two different regimes of the Purcell effect can be realized. When disorder is weak, the enhancement of spontaneous emission occurs near the edge of PBG. When the disorder increases, another regime of the Purcell coefficient establishes: edge states become smeared out and the enhancement of spontaneous emission rate occurs within PBG, due to the appearance of microcavity-like modes characterized by a high Purcell coefficient. An average maximal enhancement of spontaneous emission rate occurs near edges of PBG for weak disorder, but for specific configurations of disordered photonic crystal, maximal enhancement of spontaneous emission rate (potentially accompanied by random lasing) happens near the centre due to appearance of localized state with a high quality factor. For the PBG edges, mean value of Purcell coefficient is falling with the increase of disorder, and its standard deviation demonstrates a non-monotonic behaviour characterized by a maximum. The ratio of the standard deviation and the mean value (that is a measure of chaos in emission characteristics) demonstrates a monotonic increase with increasing disorder.

For the PBG centre, both the standard deviation and the mean value of the Purcell coefficient demonstrate a monotonic growth with disorder, while the measure of chaos in emission properties (i.e. the ratio of standard deviation and the mean value of the Purcell coefficient) demonstrates a non-monotonic dependence having a maximum.

**Acknowledgements**
The work was supported by Russian Science Foundation grant № 16-12-10503.

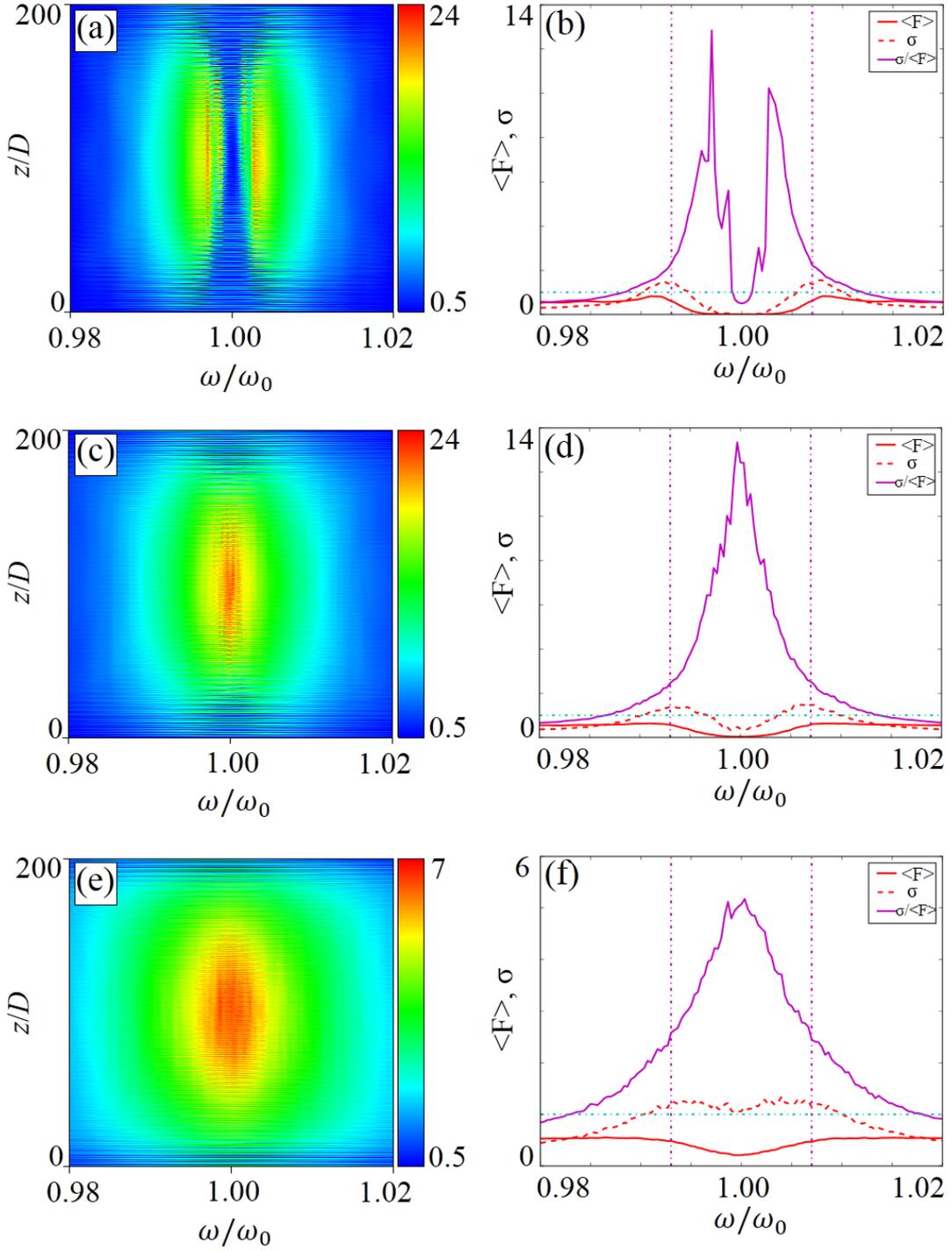

Figure 5. On the left: dependence of the ratio of the standard deviation to the averaged modal Purcell factor on the frequency and position of the dipole source placed inside the disordered structure, and averaged over an ensemble of $10^4$ structures with $\delta=0.07$ (a) 0.1 (c) and 0.15 (e). On the right: solid red lines show the dependence of the modal Purcell factor averaged over the dipole positions (in the central part of the structure [20..40] μm). Dashed lines show the standard deviation σ for the averaged Purcell factor, and the purple solid lines show the ratio of the standard deviation to the averaged modal Purcell factor, the disorder fluctuation parameter $\delta=0.07$ (b) 0.1 (d) and 0.15 (f). Dashed vertical lines indicate the edge states and the PBG of the ideal structure.

# Appendix 1.
# Calculation of Purcell coefficient using S-quantization formalism.

Solution of a wave equation for electromagnetic field in infinite uniform media with refractive index $n$

$$\nabla \times \nabla \times \boldsymbol{E} = n^2 \left(\frac{\omega}{c}\right)^2 \boldsymbol{E} \qquad (A1)$$

gives a continuous spectrum of eigenfrequencies of the mode $\omega$. In order to provide a quantum-mechanical description of the interaction of radiation and matter, the field should be quantized: continuous spectrum of EM modes should be replaced by a discrete one. For this purpose EM field is considered in a "quantization box" of "large" size (see figure A1) and boundary conditions (BC) are to be set on the facets of the box [43]. The natural choice is to set periodic (Born-Karman) BC

$$\begin{cases} E|_{x=0} = E|_{x=L_x} \\ \left.\frac{\partial E}{\partial x}\right|_{x=0} = \left.\frac{\partial E}{\partial x}\right|_{x=L_x} \end{cases} \qquad (A2a)$$

$$\begin{cases} E|_{y=0} = E|_{x=L_y} \\ \left.\frac{\partial E}{\partial y}\right|_{y=0} = \left.\frac{\partial E}{\partial y}\right|_{y=L_y} \end{cases} \qquad (A2b)$$

$$\begin{cases} E|_{z=0} = E|_{z=L_z} \\ \left.\frac{\partial E}{\partial z}\right|_{z=0} = \left.\frac{\partial E}{\partial z}\right|_{z=z} \end{cases} \qquad (A2c)$$

Wave equations (A1) with BC (A2) can be considered as eigenvalue and eigenfunction problems and the solution of the problem is given by a discrete set wavevectors $\boldsymbol{k} = (k_x, k_y, k_z)$ obeying

$$k_{x,y,z} = \pm 2\pi n N_{x,y,z}/L_{x,y,z}, \qquad (A3)$$

where $N_{x,y,z}$ are integers; corresponding eigenfunctions have the form of propagating planewaves

$$E = E_0 \exp(i(k_x x + k_y y + k_z z)) \quad (A4)$$

We should note, that the same set of eigenvalues of the wave vector is provided by equating eigenvalues of the transfer matrix $\widehat{M}$ along each direction $x$, $y$, and $z$ through the quantization box to unity, since the matrix along any particular direction (for example direction $z$) in the uniform media has a form

$$\widehat{M}_z = \begin{pmatrix} \exp(ik_z z) & 0 \\ 0 & \exp(-ik_z z) \end{pmatrix} \qquad (A5)$$

and its eigenvalues are equal to $\exp(\pm i k_z z)$.

An analysis provided above leads to an expression for the density of states in K-space

$$\rho_k = \frac{dN}{dk_x dk_y dk_z} = \frac{n^3 V}{(2\pi)^3} \qquad (A6)$$

where $V = L_x L_y L_z$ is the volume of quantization box. Then, an expression for the density of states in respect energy reads

$$\rho = \frac{dN}{d(\hbar\omega)} = \frac{dN}{dK}\frac{dK}{d(\hbar\omega)} = \frac{dN}{dK}\frac{dK}{d(\hbar\omega)} = \frac{n^3 \omega^2}{\pi^2 c^3}\frac{V}{\hbar} \qquad (A7)$$

Each EM mode can be considered as a quantum oscillator with an energy $\hbar\omega/2$, and this energy should be associated with an integral of the density of EM energy of the mode over quantization box [44]

$$\frac{1}{4\pi}\int_V n^2 E^2 d^3\boldsymbol{r} = \hbar\omega/2. \quad (A8)$$

In the case of uniform media and periodic BC, the eigenmode of EM field is nothing but a plane wave with spatially uniform amplitude, and the amplitude of the electric field for the quantized EM mode can be obtained,

$$E_0 = \frac{1}{n}\sqrt{2\pi\hbar\omega/V} \quad (A9)$$

which allows to obtain a probability of spontaneous emission $W$ for the quantum transition characterized by dipole moment $\boldsymbol{d} = e\boldsymbol{r} = er\,(\cos\varphi_d \sin\theta_d, \sin\varphi_d \sin\theta_d, \cos\theta_d)$ using Fermi golden rule:

$$W = \frac{2\pi}{\hbar}|\langle f|\boldsymbol{Ed}|i\rangle|^2 \rho \quad (A10)$$

where $\boldsymbol{E}$ is the electric field of the mode. Eq. (A6) and (A8) allow to obtain Fermi golden rule in the form which does depend on a virtual quantization box [36]:

$$W = \alpha|\langle f|\boldsymbol{\epsilon r}|i\rangle|^2 \frac{4n\omega^3}{c^2}, \quad (A11)$$

where we introduce dimensionless function $\boldsymbol{\epsilon}$ describing spatial distribution of the electric field of the mode satisfying relation $\boldsymbol{E} = E_0\boldsymbol{\epsilon}$, where $\alpha = e^2/(\hbar c) \approx 1/137$ is the fine structure constant and $\boldsymbol{\epsilon}$ is a normalized vector describing an electric field of the EM mode. Note that function $\boldsymbol{\epsilon}$ satisfies normalization condition

$$\frac{1}{4\pi}\int_V \epsilon^2 d^3\boldsymbol{r} = 1 \quad (A12)$$

As was noted above periodic BC can be set by equating eigenvalues of the transfer matrix through uniform quantization box to unity, providing a set of eigenvalues in the form of wavevectors. When inhomogeneity is inserted into the quantum box, then wavevectors are not good quantum numbers anymore. On other hand, adequate description of inhomogeneous structure can be given by a scattering matrix, which couples the waves incident on the structure (incoming waves) and outgoing waves.

We propose the procedure of quantization of electromagnetic field, based on *equating to unity eigenvalues of scattering matrix of the system*, or by equating incoming amplitudes and outgoing amplitudes.

Now we will define the quantization procedure in detail. Let us consider a quantization box with layered structure within, as shown in figure 1b.

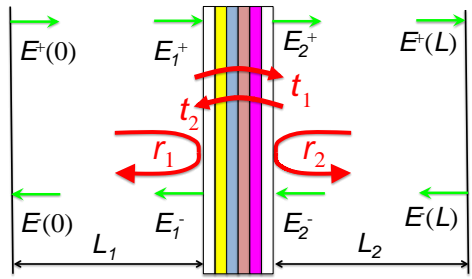

Figure A1. Inhomogeneous structure in the quantization box.

The normal to the interfaces of the layers is parallel to the $Oz$ axis, and the distances from the left and the right facet of the quantization box to layered structures are $L_1$ и $L_2$, as shown in figure A1. In the case of such layered structure, it is convenient to consider mode of electromagnetic field with specific angular frequency $\omega$ in the form

$$E_{K_x,K_y}(x,y,z) = E(z)\exp(iK_x x)\exp(iK_y y) \qquad (A13a)$$

where the lateral components of the wavevector $K_x$ and $K_y$ relate to direction of propagation of the waves in empty parts of quantization box via relations

$$K_x = \frac{\omega}{c}\sin\theta\cos\varphi \qquad (A13b)$$

and

$$K_y = \frac{\omega}{c}\sin\theta\sin\varphi. \qquad (A13c)$$

In the case of TE polarization electric field of the wave has the component $E_y$ only, while for TM polarization there are components $E_x$ and $E_z$.

In each layer of the structure, spatial dependence of electric field along $z$-axis is defined as superposition of the waves propagating in opposite directions along $z$–axis, and in the subsequent discussion we denote the wave with positive $K_z$ with upper index "+ ", for negative $K_z$ we will use upper index " - ".

We denote amplitudes of the waves incident on the left and right facets of the quantization box as $E^+(0)$ and $E^-(L)$, and amplitudes of the waves outgoing from right and left boundaries as $E^+(L)$ and $E^-(0)$.

Amplitudes of the waves on left and right facets of the quantization box are coupled by relation

$$\begin{pmatrix} \lambda_2^* E^+_{K_x,K_y}(L) \\ \lambda_1^* E^-_{K_x,K_y}(0) \end{pmatrix} = \begin{pmatrix} t_1 & r_1 \\ r_2 & t_2 \end{pmatrix} \begin{pmatrix} \lambda_1 E^+_{K_x,K_y}(0) \\ \lambda_2 E^-_{K_x,K_y}(L) \end{pmatrix} \qquad (A14)$$

where $r_1$ and $r_2$ are the amplitude reflection coefficients of layered structure for the waves incident from the left and right sides respectively, $t_1$ and $t_2$ are the corresponding amplitudes of transmission coefficient of layered structure, and the phases gained by waves propagating from the facets of quantization boxes to layered structures are given by $\lambda_{1,2} = \exp(iK_z L_{1,2})$. It follows that amplitudes of incoming waves $[E^+_{K_x,K_y}(0), \; E^-_{K_x,K_y}(L)]$ are coupled with amplitudes of outgoing waves $[E^+_{K_x,K_y}(L), \; E^-_{K_x,K_y}(0)]$ by scattering matrix $\hat{S}$

$$\begin{pmatrix} E^+_{K_x,K_y}(L) \\ E^-_{K_x,K_y}(0) \end{pmatrix} = \hat{S} \begin{pmatrix} E^+_{K_x,K_y}(0) \\ E^-_{K_x,K_y}(L) \end{pmatrix} \qquad (A15)$$

and $\hat{S}$ reads

$$\hat{S} = \begin{pmatrix} \lambda_1\lambda_2 t_1 & \lambda_2^2 r_2 \\ \lambda_1^2 r_1 & \lambda_1\lambda_2 t_2 \end{pmatrix}. \qquad (A16)$$

Eigenvalues of $\hat{S}$ matrix reads

$$\beta_{1,2} = \lambda_1\lambda_2\left(\frac{t_1+t_2}{2} \pm \sqrt{\left(\frac{t_1-t_2}{2}\right)^2 + r_1 r_2}\right) \qquad (A17a)$$

In an important case of a non-absorbing system or for any system possessing centre of symmetry eigenvalues have simple form

$$\beta^{(1,2)} = \lambda_1\lambda_2(t \pm \sqrt{r_1 r_2}), \qquad \text{(A17b)}$$

and related eigenvectors are

$$B^{(1,2)} = \left[1, -\frac{\lambda_1}{\lambda_2}\left(\frac{t_2-t_1}{2r_2} \pm \sqrt{\left(\frac{t_2-t_1}{2r_2}\right)^2 + \frac{r_1}{r_2}}\right)\right] \qquad \text{(A18a)}$$

for non-absorbing or centrally symmetric system eigen-vectors reads

$$B^{(1,2)} = \left[1, \; \pm(\lambda_1/\lambda_2)\sqrt{r_1/r_2}\,\right]. \qquad \text{(A18b)}$$

It is clear, that for a symmetric quantization box (when $L_1 = L_2$) the eigen-vectors depends only on the properties of the inhomogeneity, but NOT on the size of quantization box.

Periodic BC imply equating the field on the opposite sides of the quantization box, which is equivalent to equating eigenvalues of the transfer matrix to unity. Here we provide quantization of the field using different BC: we equate "incoming" and "outgoing" fields, what means equating the eigenvalues of the scattering matrix $\hat{S}$ to unity:

$$\beta^{(1,2)} = 1 \qquad \text{(A19)}$$

Solution of eq. (A19) in respect to frequency thus gives the spectrum of eigenfrequencies. Using the set of quantum numbers, one can obtain the eigenvectors $B^{(1)}$ and $B^{(2)}$, and calculate the field profile of the mode using the transfer matrix method. The components of the eigenvectors $B^{(1,2)}$ are the complex amplitudes of the fields incident on the edges of the box, and the field of the mode is the superposition of the fields, excited by waves incident on the structure from opposite directions, and corresponding spatial profiles of the electric field described by the functions $\tilde{\epsilon}^{(1,2)}$.

Similar to the case of periodic BC, we can consider the mode obtained using S-quantization as an elementary quantum oscillator and normalize it using equation (A8). The field of the mode should be normalized according to eq. (A12). We denote BC (A19) as S–conditions, and the procedure of quantization described above as S-quantization.

In the case of uniform media, BC given by eq. (A19) is nothing but periodic BC. At the same time, the modes defined by eigenvectors $B^{(1)}$ and $B^{(2)}$ with filed distribution described by functions $\epsilon^{(1)}$ and $\epsilon^{(2)}$ respectively, will not be plane waves, propagating in opposite directions, but will be standing waves of equal amplitude, shifted by the quarter of a wavelength.

In non-absorbing media eigenfrequencies defined by eq. (A19) are real, which reflects equity of incoming and outgoing fluxes. Eq. (A17) can be rewritten in the form

$$\beta^{(1,2)} = \exp(iK_z L + \alpha) \qquad \text{(A20)}$$

where $\alpha$ is a phase, defined by reflection and transmission coefficients of layered structure and is depending on frequency of the light. When the size of the quantization box is large enough in respect to layered structure, $K_z L$ varies much faster then $\alpha$ with increasing frequency, and one-dimentional density of states in $K$-space reads $\frac{dN}{dK_z} = L/(2\pi)$, as in the case of uniform media, and 3D density of states is given by eq. (A6).

In non-uniform media, spatial envelope function of the electric field of the mode is not constant, and the probability of spontaneous emission is defined by the magnitude of an electric field of the mode at the position of the emitting dipole. For each eigenvalue $\beta^{(1,2)}$ the spatial profile of the field of EM mode is a superposition of the fields excited by the two waves incident from left and right sides of the structure with the frequencies defined by eq. (A19), and the amplitudes of these two waves are coupled by eq. (A18). As usual, each mode should be considered as elementary quantum oscillator, and normalized using eq. (A8).

Let us consider the situation when the K-vector of light in an empty quantization box is within the light cone, i.e. $K_z < \omega/c$. In this case light can leak from the structure into the quantization box. If the size of the quantization box goes to infinity, then the contribution of the layered structure to value of integral (A8) will be negligible, and the integral (A8) will be equal to the contribution given by the wave in empty parts of quantization box. Thus, the amplitude of electric field of EM mode, normalized using eq. (A8) incident on empty quantization box, and incident on quantization box with layered structure, will be equal.

Since density of the states provided by S-quantization is the same density of states as setting periodic BC, for the specific EM mode probability of spontaneous emission given by eq. (A10) for the dipole in layered structure will be defined by modification of the amplitude of electric field vector $\tilde{\epsilon}$ in layered structure. *This modification (spatial profile of the field within layered structure) does not depend on the size of the left and right empty parts of quantization box, and is defined only by reflection coefficients $r_1$ and $r_2$ and transmission coefficient t of the layered structures. Thus, the size of empty part of quantization box can be reduced to zero.*

An approach based on modification of the spatial profile of the modes in microcavities has been used by De Martini [31], though the use of the periodic BC limits an applicability of results obtained in this work. An approach used in [31], corresponds to the use of only "symmetric" eigenvector $B^{(1)}$, while the mode corresponding to "antisymmetric" eigenvector $B^{(2)}$ is missed in [31]. However, if an emitter is placed at the centre of a symmetric structure (as has been done in [31]) absence of $B^{(2)}$ does not affect the validity of the results, since the value of the mode field corresponding to $B^{(2)}$ is zero in centre of the symmetric structure. If the dipole is placed into arbitrary place in the structure without specific symmetry, the modes corresponding to both $B^{(1)}$ and $B^{(2)}$ should be taken into account.

For a development of the formalism it is convenient to relate components of vector $\tilde{\epsilon}^{(1,2)}$ describing electric field in layered structure to the components of vector $\epsilon^{(1,2)}$ for uniform medium, via coefficients X, Y and Z as specified below. For TE mode

$$\tilde{\epsilon}^{(1,2)} = \begin{pmatrix} 0 & \tilde{\epsilon}_y^{(1,2)} & 0 \end{pmatrix} = \begin{pmatrix} 0 & Y^{(1,2)}\epsilon_y^{(1,2)} & 0 \end{pmatrix} \qquad (A21)$$

while for TM mode

$$\tilde{\epsilon}^{(1,2)} = \begin{pmatrix} \tilde{\epsilon}_x^{(1,2)} & 0 & \tilde{\epsilon}_z^{(1,2)} \end{pmatrix} = \begin{pmatrix} X^{(1,2)}\epsilon_x^{(1,2)} & 0 & Z^{(1,2)}\epsilon_z^{(1,2)} \end{pmatrix} \qquad (A22)$$

It is also convenient to define the *modal Purcell factor* for a specific mode characterized by direction of propagation defined by the polar angle $\theta$ of wave in free space as a ratio of probability of spontaneous emission for this mode to probability of spontaneous emission in the free space, when dipole is parallel to the field of the mode:

$$F_\theta^{(TE)} = \frac{|\langle f|\tilde{\epsilon}r|i\rangle|^2}{|\langle f|\epsilon r|i\rangle|^2} \qquad (A23)$$

Such definition of modal Purcell factor will be convenient for the subsequent analysis of the Purcell effect in the case of waveguide modes.

Similarly, the dot product in eq. (A23) for TE modes reads

$$\tilde{\boldsymbol{\epsilon}}^{(1,2)}\boldsymbol{r} = \tilde{\epsilon}_y^{(1,2)} r_y = Y^{(1,2)} \epsilon_y^{(1,2)} r_y \qquad (A24)$$

and for TM mode

$$\tilde{\boldsymbol{\epsilon}}^{(1,2)}\boldsymbol{r} = \tilde{\epsilon}_x^{(1,2)} r_x + \tilde{\epsilon}_z^{(1,2)} r_z = X^{(1,2)} \epsilon_x^{(1,2)} r_x + Z^{(1,2)} \epsilon_z^{(1,2)} r_z \qquad (A25)$$

Therefore, the Purcell factor for specific TE mode characterized by emission angle $\theta$ is

$$F_\theta^{(TE)} = \sum_{i=1,2} |Y^{(i)}|^2 (r_y/r)^2 = \sum_{i=1,2} |Y^{(i)}|^2 \sin^2 \varphi_d \sin^2 \theta_d \qquad (A26)$$

while for TM mode

$$F_\theta^{(TM)} = \sum_{i=1,2} \left| X^{(i)} \frac{\epsilon_x^{(i)}}{|\epsilon^{(i)}|} \frac{r_x}{r} + Z^{(i)} \frac{\epsilon_z^{(i)}}{|\epsilon^{(i)}|} \frac{r_z}{r} \right|^2 =$$

$$= \sum_{i=1,2} |X^{(i)} \cos\theta \cos\varphi_d \sin\theta_d + Z^{(i)} \sin\theta \cos\theta_d|^2 \qquad (A27)$$

In the case of TE polarization, for the dipole oriented along y-axis, the modal Purcell factor is nothing but

$$F_\theta^{(TE)} = \sum_{i=1,2} |Y^{(i)}|^2 \qquad (A28a)$$

For the dipole, oriented along axis $Ox$ the Purcell factor for TM modes reads

$$F_\theta^{(TM)} = \sum_{i=1,2} |X^{(i)}|^2 \cos^2 \theta \qquad (A28b)$$

and for the orientation of dipole along $Oz$ axis

$$F_\theta^{(TM)} = \sum_{i=1,2} |Z^{(i)}|^2 \sin^2 \theta \qquad (A28c)$$

Thus the quantities $X$, $Y$, and $Z$ define the probability of spontaneous emission in layered structure within the light cone.

Since the size of the quantization box does not influence the components of eigenvectors $B^{(1,2)}$ we can exclude the quantization box from consideration. Thus, for the construction of eigen-vectors and the functions $\tilde{\boldsymbol{\epsilon}}^{(1,2)}$, one can take the values of the reflection and transmission coefficients of inhomogeneity at its interfaces.